\documentclass[12pt]{article}

\def\mytitle#1{\setcounter{equation}{0}
\setcounter{footnote}{0}
\begin{flushleft}\Large\textbf{#1}\end{flushleft}
\vspace{0.27cm}}
\def\myname#1{\leftline{{\large #1}}\vspace{-0.13cm}}
\def\myplace#1#2{\small\begin{flushleft}\textit{#1}\\
\texttt{#2}\end{flushleft}}

\def\myclassification#1{\small\noindent
Pacs no : 98.80.Es, 98.80Cq, 04.20.-q, 04.60.Cf
       #1\vspace{0.5cm}}
\usepackage{graphicx}
\begin{document}

\mytitle{Will there be again a transition from acceleration to deceleration in course of the dark energy evolution of the universe?}

\vskip0.2cm \myname{Supriya Pan\footnote{pansupriya051088@gmail.com}}
\vskip0.2cm \myname{Subenoy Chakraborty\footnote{schakraborty@math.jdvu.ac.in}}

\myplace{Department of Mathematics, Jadavpur University, Kolkata-700 032, India.}{}

\begin{abstract}

In this work we consider the evolution of the interactive dark fluids in the background of homogeneous and isotropic FRW model of the universe. The dark fluids consist of a warm dark matter and a dark energy and both are described as perfect fluid with barotropic equation of state. The dark species interact non-gravitationally through an additional term in the energy conservation equations. An autonomous system is formed in the energy density spaces and fixed points are analyzed. A general expression for the deceleration parameter has been obtained and it is possible to have more than one zero of the deceleration parameter. Finally, vanishing of the deceleration parameter has been examined with some examples.\\

\end{abstract}
\myclassification{}\\
\section{Introduction}

Recent observational data coming from Supernove Ia (SNe Ia) [1], Large Scale Structure formation (LSS) [2], Cosmic Microwave Background Radiation (CMBR) [3], Baryon Acoustic Oscillation (BAO) [4] and Weak Lensing [5] put a big question mark to the standard cosmology. As the gravity of both baryonic (ordinary) matter and radiation is attractive in nature, so standard cosmology predicts a deceleration of the universe while observationally the universe is currently going through an accelerated expansion. To overcome this challenging and intriguing problem in cosmology people are trying to modify the standard cosmology in two ways: {\bf (I)} by incorporating a new form of matter called dark energy (DE) which has marginally dominant negative pressure component that drives the present acceleration of the universe [6]. {\bf(II)} Secondly, the doubt about the applicibality of the General Relativity to describe the universe as a whole, results alternative theories of gravity. Also alternative theories of gravity are proposed to resolve the puzzles of theoretical and experimental gravity [7]. These include f(R) gravity theory [8], theories with other curvature invariants [9], Coupling the Ricci scalar to a scalar field [10] by introducing a vector field contribution [11], or by gravity theory in higher dimensional space-times [12]. In these theories to produce cosmic acceleration one does not require any exotic matter (i.e, DE) rather it is provided by the extra terms in the usual Friedmann equations on large scales [13].\\
Due to lack of any definite conclusion, various cosmological models have been proposed to match with recent observed data. $\Lambda$CDM model is the simplest one in this series. Although the model predicts cosmic acceleration as well as a reasonable agreement with observational data, but there are some embarrassing issues related to this model namely cosmological constant problem [14] (the huge discrepency between the observed value of the cosmological constant and the one predicted in quantum field theory), coincidence problem [15] (although generically small, but the cosmological constant happens to be exactly of the value required to become dominant at the present epoch) and recently, it was shown that the $\Lambda$CDM model may also suffer the age problem [16].\\
To overcome these puzzles, it is natural to persue alternative possibilities to explain the mystery of dark energy. In the literature it is assumed that the energy of the vacuum is zero (by some unknown cancellation mechanism) and in its place a dark energy components having dynamically variable equation of state is considered. Over the last decade, various dark energy models have been proposed namely, Quintessence [17], Phantom [18], K-esssence [19], Tachyon [20], Quintom [21], chaplygin gas [22], Holographic DE [23], the new age graphic [24], the Ricci DE [25] and so on. Motivated by the supersymmetric field theories and String/ M theory these DE models are represented by some effective scalar fields.\\
Further, though both the dark components (DE and DM) evolve differently, but cosmic observations predict their energy densities of the same order today. To overcome this Coincidence problem [26], both the dark components are assumed to interact non-gravitationally [27] through an additional term in the conservation equations. Also in the perspective of the present data the DE should be chosen such that there is a smooth transition across the phantom barrier from the above in near past [28].\\
In the present work, we propose a phenomenological cosmological model where the two dark species interact among themselves. The equation of state parameter for warm dark matter is chosen to be positive but gradually decreases while the DE equation of state parameter is negative throughout the evolution but otherwise unrestricted. The model describes the evolution of the universe from an initial dense radiation era (or a barotropic perfect fluid) when DM dominates over DE. Subsequently, DE takes an upper hand and the universe gradually enters into the accelerating phase. We mainly focus our attention how the evolution of the universe and its cosmic acceleration depend on the evolving equation of state parameters of the two dark specieses. we shall also examine (by examples) whether the accelerated expansion will continue forever or it is possible to have an ever expanding DE-dominated universe which passes through from deceleration to acceleration and vice versa.\\
The paper is organized as follows: Basic equations for the phenomenological model has been presented in section 2. For two specific choices of the interaction term, the energy conservation equations are reduced to an autonomous system and critical points are analyzed in section 3. Section 4 presented a general prescription for the deceleration parameter and in section 5, it has been examined whether more than one transition is possible or not by some examples. Finally, in section 6, there is a brief discussion about the results obtained in this work.

\section{Basic equations:}
We start with homogeneous and isotropic FRW model of the universe filled with dark fluids- a warm dark matter of energy density $\rho_m$ and a dark energy component described by the dark energy density $\rho_d$. As we are interested with the late time cosmic evolution so we have neglected other matter components namely radiation, baryons etc. For simplicity, both the dark fluids are assumed to have barotropic equation of state namely,
\begin{equation}
p_m =\omega_m \rho_m~~~~~and~~~~~~
p_d =\omega_d \rho_d
\end{equation}
where $\omega_m$ ($> 0$) and $\omega_d$ ($<0$) are in general chosen as variables. So the Einstein field equations take the form
\begin{equation}
3(H^2 +\frac{\kappa}{a^2}) =\rho_m+\rho_d
\end{equation}
and
\begin{equation}
\dot{H}-\frac{\kappa}{a^2} =-\frac{1}{2}[\rho_m(1+\omega_d)+\rho_d(1+\omega_d)]
\end{equation}\\
If we assume that the two dark components interact among themselves, then the continuity equations are of the form
\begin{equation}
\dot{\rho_m} +3H\rho_m(1+\omega_m) =Q
\end{equation}
and
\begin{equation}
\dot{\rho_d} +3H\rho_d(1+\omega_d) =-Q
\end{equation}\\
where Q stands for the interaction between the dark sectors. For the time being we do not use any specific choice for Q, only assume that Q does not change its sign during the cosmic evolution. Now the above continuity equations can be written in the non-interaction form with effective equation of state parameters as 
\begin{equation}
\omega_{me} =\omega_m-\frac{Q}{3H\rho_m} ~~~~~~~~~and~~~~~~~~~~~~  
\omega_{de} =\omega_d+\frac{Q}{3H\rho_d}
\end{equation}\\
As usual, if we assume that the energy is transferred from dark energy (DE) to dark matter (DM), i.e., $Q>0$, then we have the following observations:\\

{\bf(a)} For cold DM (i.e., $\omega_m = 0$) we have effectively some kind of exotic DM (i.e., -ve equation of state parameter) in the presence of interaction.\\

{\bf(b)} For warm DM (i.e., $\omega_m > 0$) it is likely to have a possible change of sign of the effective equation of state during cosmic evolution depending on the nature of the interaction and the strength of the coupling constant appeaaring in Q.\\

{\bf(c)} If the DE is chosen as the cosmological constant (i.e., $\omega_d = -1$) then the effective DE will behave as a quintessence field (i.e., $\omega_{de} > -1$). Also the effective DE fluid may be quintessence era even we choose phantom DE (i.e., $\omega_d <-1$).\\
We now introduce the dimensionless parameter r = $\frac{\rho_m}{\rho_d}$, which is also termed as the coincidence parameter. Using the energy conservation equations (4) and (5), the evolution equation for r with respect to $\tau$ (= 3lna) is given by (after some simplification)
\begin{equation}
\frac{dr}{d\tau} =r[(\omega_d-\omega_m)+\frac{Q}{3H}\frac{(1+r)^2}{r\rho_t}]
\end{equation}\\
where $\rho_t =\rho_m+\rho_d$, is the total energy density of the dark fluids. Also using the above energy conservation equations, the evolution of $\rho_t$ is given by
\begin{equation}
\frac{d\rho_t}{d\tau} =-[1+\frac{r\omega_m+\omega_d}{1+r}]\rho_t
\end{equation}\\
The fixed points of equations (7) and (8) are given by 
\begin{equation}
r_c = -\frac{1+\omega_d}{1+\omega_m}
\end{equation}
and
\begin{equation}
\rho_c = -\frac{Q}{3H}\frac{\omega_m-\omega_d}{(1+\omega_m)(1+\omega_d)}
\end{equation}\\
where in the last equation we have used equation (9) for $r_c$. It is to be noted that as $r_c$ and $\rho_c$ are positive definite so we must have phantom DE, i.e., $\omega_d < -1$. If $\omega_m = 0$, i.e., for cold DM to have a realistic fixed point (i.e., +ve stationary energy density and coincidence parameter) there must be an exchange of energy from phantom DE to cold DM. The same is true for warm DM also. It is to be noted that the result still remains valid even for $-1 < \omega_m < 0$.\\
\section{Fixed points and stability analysis for Holographic dark energy model:}
In this section we shall choose a common model of DE which follows the holographic principle known as holographic dark energy (HDE). The effective quantum field theory determines the holographic energy density as [23]\\
$$ \rho_d= 3 c^2 M^2 _p L^{-2}$$\\
where L is an IR cut off in units $M^2 _p= 1$. It is found that [29], $L= R_E$, the radius of the event horizon gives the correct equation of state and the desired accelerating universe. Further, it should be noted that in the above expression for $\rho_d$, c is any free dimensionless parameter, may be estimated from the observational data. However, in the present work c is taken as arbitrary. So from the above equation
\begin{equation}
R_E =\frac{c}{H\sqrt{\Omega_d}} 
\end{equation}

where $\Omega_d = \frac{\rho_d}{3H^2}$ is the usual density parameter for DE.\\

Now using the conservation equation (5) and the Friedman equations (2) and (3) the evolution of the DE can be written as
\begin{equation}
\frac{\Omega^\prime_d}{\Omega_d} =(1- \Omega_d)[3\omega_m +1+\frac{2\sqrt{\Omega_d}}{c}]-\frac{Q\Omega_d}{H\rho_d}
\end{equation}\\
where, $^\prime$ stands for differentiation with respect to x = lna. The equation of state parameter for the HDE has the form
\begin{equation}
\omega_d = -\frac{1}{3}(1+2\frac{\sqrt{\Omega_d}}{c})-\frac{Q}{3H\rho_d}
\end{equation}\\
Note that $\omega_d$ does not depend explicitly on $\omega_m$. Also it should be mentioned that if $\omega_m = 0$ then with proper choice of Q, one gets back to equations (5) and (6) of reference [30] (or equations (23) and (24) of reference [31]) from equations (12) and (13) respectively. We shall analyze the evolution equations for the following two choices of the interaction term:
\begin{equation}
{\bf (a)} Q = 3Hb^2(\rho_m+\rho_d)~~~~~~~~and ~~~~~~~  {\bf (b)} Q = 3\lambda H \rho_d
\end{equation}\\
with $b^2$ and $\lambda$ as the coupling constants.\\

{\bf Case I:} {$\bf Q=3Hb^2(\rho_m+\rho_d)$}\\

This expression for the interaction term was first introduced to study coupling between a quintessence scalar field and a pressureless cold DM field [27], so that the coincidence problem can be resolved by scaling solution and the universe approaches a stationary stage (i.e., r = constant). Subsequently, this form of interaction was derived in the context of HDE with Hubble scale as the IR cut off [27] and also with event horizon as the IR cut off [31]. Now for this choice of Q and using the equation of state parameter $\omega_d$ for HDE from equation (13), the energy conservation equations can be written as 
\begin{equation}
\rho{^\prime}_m = 3\rho_m[b^2(1+\frac{1}{r})-(1+\omega_m)] 
\end{equation}
and
\begin{equation}
\rho{^\prime}_d =-2\rho_d[1-\frac{1}{c\sqrt{1+r}}]
\end{equation}\\
If $\omega_m$ is assumed to be constant then equations (15) and (16) form an autonomous system in the ($\rho_d$, $\rho_m$) plane. Hence at the fixed points the coincidence parameter has the constant value
\begin{equation}
r_0 = \frac{1}{c^2} -1
\end{equation}\\
which implies $c^2 < 1$. Also $\omega_m$ corresponding to the fixed points has the value
\begin{equation} 
\omega_m = b^2(1+\frac{1}{r_0}) -1 = \frac{b^2}{1-c^2} -1
\end{equation}\\
As $\omega_m > 0$, so $c^2$ is restricted by the relation
\begin{equation}
1-b^2 < c^2 < 1
\end{equation}\\
However, in the plane ($\rho_d$, $\rho_m$) the fixed points lie on a straight line with positive slope ($\frac{1}{c^2} -1$). The phase portrait has been presented in Figure 1. In particular, if the DM is chosen in the form of Dust (i.e., Cold DM), then we have $b^2 = 1 -c^2$, $r_0 = \frac{b^2}{c^2}$ and hence the coincidence problem will have a solution by proper choice of the coupling parameters.
$$~~~~~~~~~~~~~~~~~~~~~~~~~$$\includegraphics[height= 4 in, width= 6 in]{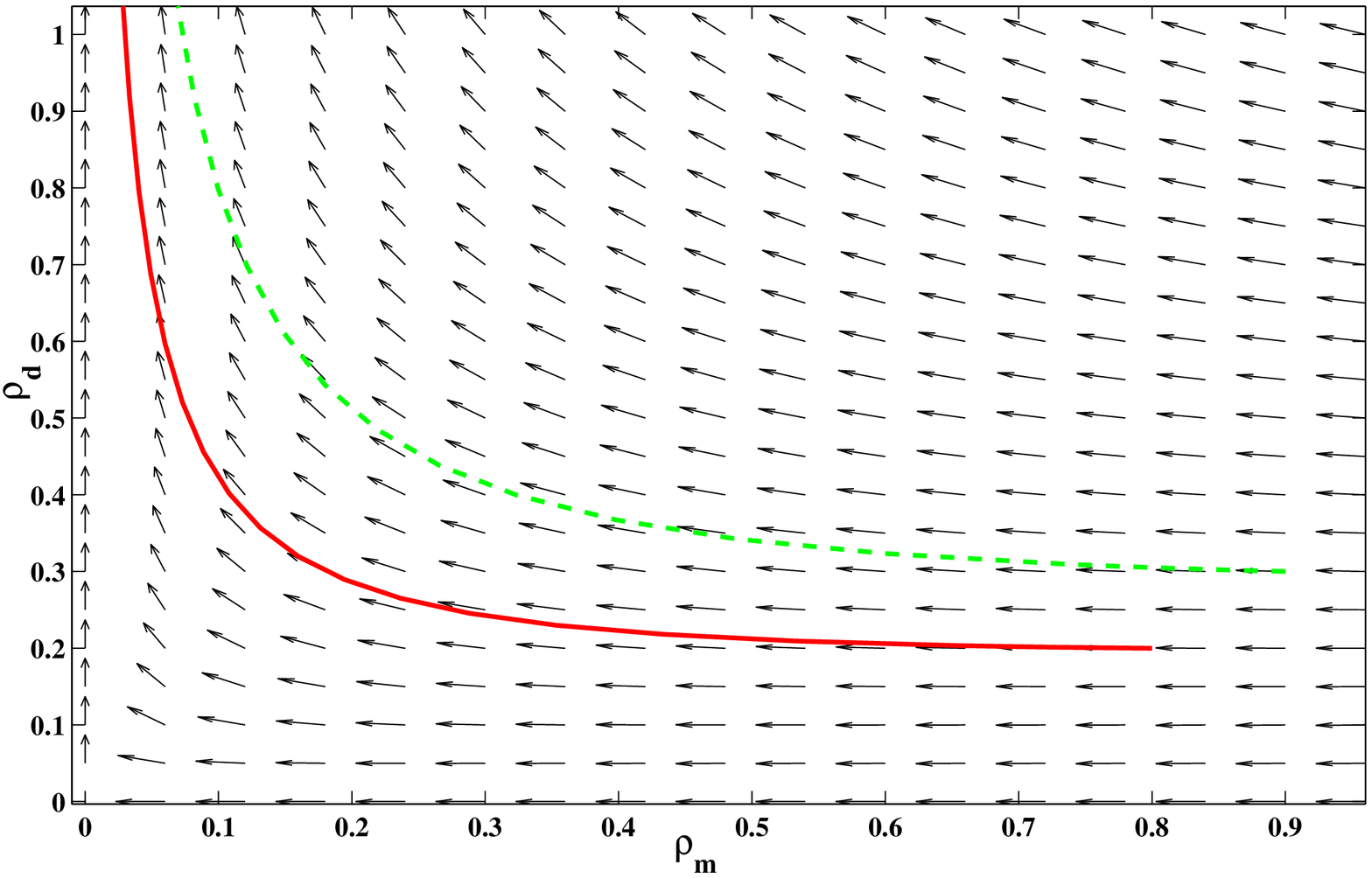}\\
Figure 1: The figure shows the phase portrait around the line of fixed points, where for the solid and dashed curves the initial values of $(\rho_m, \rho_d)$ are (0.8, 0.2) and (0.9, 0.3) respectively. For both the curves we have taken b = 0.01, $\omega_m$ = 0.3 and c = 0.4.\\\\

{\bf Case II:} {$\bf Q = 3\lambda H \rho_d$}\\

This choice of interaction term appears to be phenomenological but it is compatible with observations [32] like Sn Ia, CMBR, LSS, H(z), age constraints and recently in galactic clusters observations. Here $\lambda$ is a small dimensionless positive parameter and factor '3' is chosen for simplicity of calculations.\\

In this case, the energy conservation equations have the explicit form\\

$\rho{^\prime}_m = 3\rho_m[\frac{\lambda}{r} -(1+\omega_m)]$~~~~~~~~~~~~~~~~~~~~~~~~~~~~~~~~~
and~~~~~~~~~~~~~~~~~~
$\rho{^\prime}_d = -2\rho_d[1-\frac{1}{c\sqrt{1+r}}]$\\

As before they form an autonomous system for constant $\omega_m$ in the ($\rho_d$, $\rho_m$) plane and the fixed points are characterized by\\

$\omega_m = \frac{\lambda}{r_0} -1$,~~~~~~~~~~~~~~~~~~~~~~$r_0 = \frac{1}{c^2} -1$\\
As $\omega_m \geq 0$, so the parameters are restricted by the relation\\

$\frac{1}{1+\lambda^2} \leq c^2 < 1$\\

where the equality sign holds for cold DM. Similar to the previous case we have a line of fixed points with positive slope $\frac{1-c^2}{c^2}$ in ($\rho_d$, $\rho_m$) plane as shown in Figure 2.\\
$$~~~~~~~~~~~~~~~~~~~~~~~~~~$$\includegraphics[height= 4 in, width= 6 in]{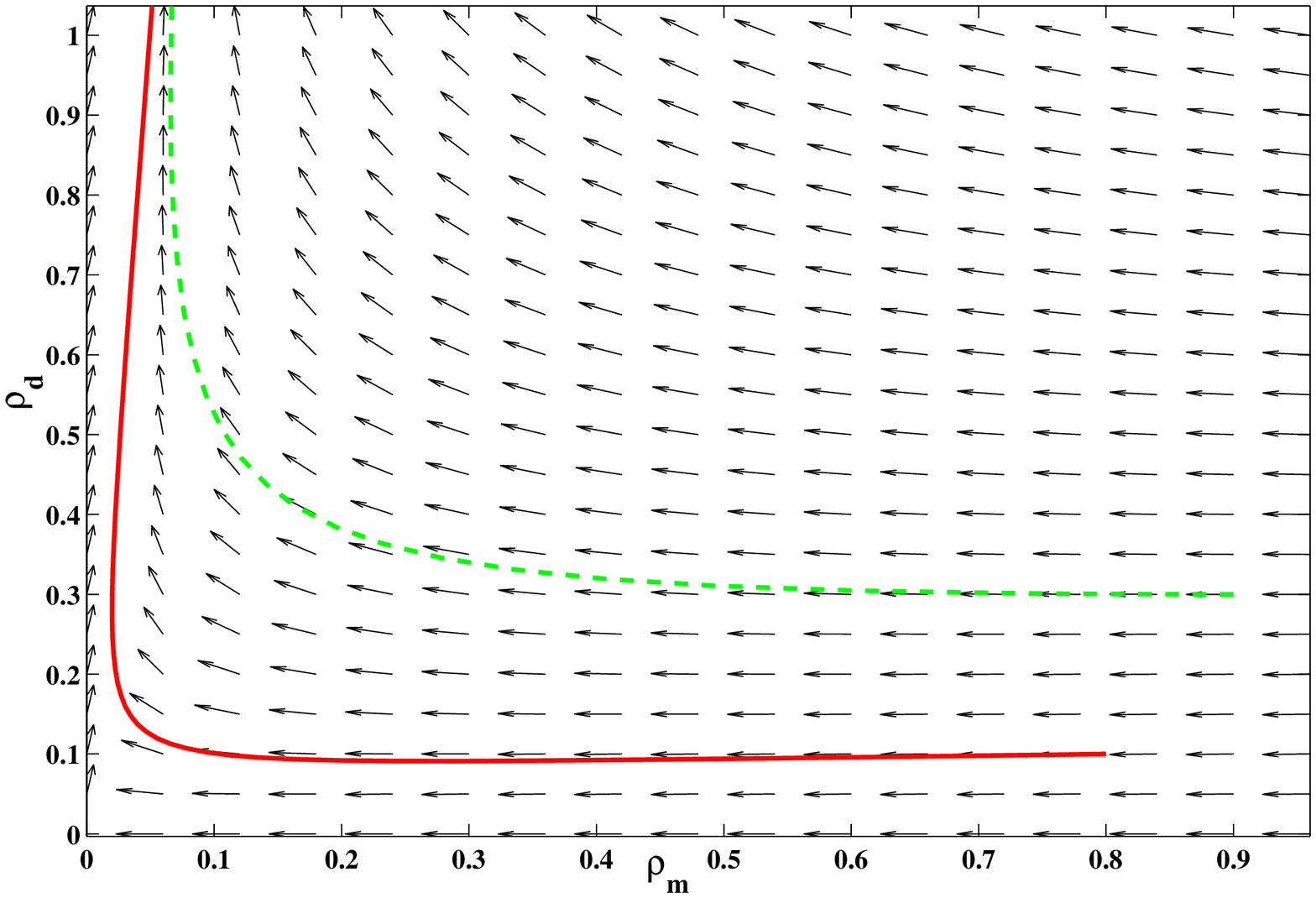}\\
Figure 2: The figure gives an idea of the phase portrait around the line of critical points where for the solid and dashed curves the initial values of $(\rho_m, \rho_d)$ are (0.8, 0.1) and (0.9, 0.3) respectively. For both the curves we have taken $\lambda$ = 0.1, $\omega_m$ = 0.4 and c = 0.5.\\

In the above discussion for both the cases the self-autonomous system can be written as $\rho^\prime = \phi(\rho)$ with $\rho = (\rho_m, \rho_d)$. To determine the stability criteria around the fixed points we expand the $\rho$ around $\rho_0$ (critical point), i.e., $\rho = \rho_0 + x$, where the perturbation of the variables x satisfies (upto first order) $x^\prime = Mx$. The linearized perturbation matrix M characterizes the fixed points and the stability criteria is determined by Tr M $< 0$ and det M$> 0$. However, in the present problem we have det M= 0 (in both the cases), so the fixed points are degenerate equilibrium points. Further, the eigenvalues of the linearized matrix M  are 0 and Tr M, so the fixed points are non-hyperbolic in nature and it is not possible to obtain any conclusive information about the stability from linearization- one needs Normal forms calculations [33] or numerical experimentations.
\section{A general prescription for deceleration parameter:}
Introducing the density parameters $\Omega_m = \frac{\rho_m}{3H^2}$ and $\Omega_d = \frac{\rho_d}{3H^2}$ and the deceleration parameter $q = -(1+\frac{\dot{H}}{H^2})$, the field equations (2) and (3) can be written as 
\begin{equation} 
\Omega_m +\Omega_d = 1+\Omega_{\kappa}
\end{equation}
and
\begin{equation}
q = \frac{1}{2}(1+\Omega_{\kappa})+\frac{3}{2}(\Omega_d\omega_d+\Omega_m\omega_m)
\end{equation}\\
with $\Omega_{\kappa} = \frac{\kappa/a^2}{H^2}$.\\

Now solving for $\Omega_m$ and $\Omega_d$ we obtain
\begin{equation}
\Omega_m = [2q-(1+3\omega_d)(1+\Omega_{\kappa})]/3(\omega_m-\omega_d)
\end{equation}
and
\begin{equation}
\Omega_d = [(1+\omega_m)(1+\Omega_{\kappa})-2q]/3(\omega_m-\omega_d)
\end{equation}\\
Hence the positivity of the density parameters restrict the deceleration parameter as 
\begin{equation}
\frac{1+3\omega_d}{2}(1+\Omega_{\kappa}) \leq q \leq \frac{1+3\omega_m}{2}(1+\Omega_{\kappa})
\end{equation}\\
which for the $\Lambda$CDM model (i.e., $\omega_m = 0$, $\omega_d = -1$, $\kappa = 0$ ) simplifies to 
$-1\leq q \leq \frac{1}{2}$ and describes the whole evolution from dust era to $\Lambda$CDM. On the otherhand, during DE dominance (i.e., $\frac{1}{2} \leq \Omega_d \leq 1$, $0 < \Omega_m \leq \frac{1}{2}$, with $\Omega_{\kappa} =0$), we have from (21)
\begin{equation}
\frac{1+3\omega_d}{2} \leq q  \leq \frac{1}{2} +\frac{3}{4}(\omega_m+\omega_d)
\end{equation}\\ 
which on the phantom barrier (i.e., at the extreme DE era) simplifies to $-1 \leq q \leq -\frac{1}{4}$.\\
If we consider the system of two dark fluids as a single fluid with energy density $\rho_t = \rho_m+\rho_d$, then the effective equation of state of the single fluid is 
\begin{equation}
\omega_t = \frac{\Omega_m\omega_m+\Omega_d\omega_d}{1+\Omega_{\kappa}}
\end{equation}\\
which does not depend on the interaction term. Hence the deceleration parameter simplifies to 
\begin{equation} 
q = \frac{1}{2}(1+3\omega_t)(1+\Omega_{\kappa})
\end{equation}\\
which is the usual definition of the deceleration parameter for a single fluid.\\

Further, using the effective equation of state parameters from equations (6), the continuity equations (4) and (5) can be integrated to give
\begin{equation}
\rho_m =\frac{\rho_0}{u^3}\alpha(u)~~~~~~~~~~~~and~~~~~~~~~~
\rho_d = \frac{\rho_0}{u^3}\beta(u)
\end{equation}\\
where u is a dimensionless variable defined as [34]
\begin{equation}
u = \frac{a}{a_*},
\end{equation}\\
$a_* = a (t_*)$ is the value of the scale factor at the instant $t_*$ when energy densities of both the dark components are identical (i.e., $\rho_m(t_*) = \rho_d(t_*) = \rho_0 $(say)).\\
Here,
\begin{equation}
\alpha(u) = \exp[-3{\int_1}^u\frac{\omega_{me}(x)}{x}dx]~~~~~~~~~~~~~~~and~~~~~~~~~~~~~
\beta(u) = \exp[-3{\int_1}^u\frac{\omega_{de}(x)}{x}dx]
\end{equation}\\

For flat space, the density parameters can be written as 
\begin{equation}
\Omega_m = \frac{\alpha(u)}{\alpha(u)+\beta(u)}~~~~~~~~~~~~~~~~~~and~~~~~~~~~~~~~
\Omega_d = \frac{\beta(u)}{\alpha(u)+\beta(u)}
\end{equation}\\
and the deceleration parameter takes the form
\begin{equation}
q(u) = \frac{1}{2}+\frac{3}{2}[\frac{\alpha(u)\omega_m+\beta(u)\omega_d}{\alpha(u)+\beta(u)}]
\end{equation}\\

Now differentiating (32) with respect to u we get (after some simplification)
\begin{equation}
\frac{dq}{du} = \frac{3}{2}(\Omega_m\frac{d\omega_m}{du}+\Omega_d\frac{d\omega_d}{du})+\frac{9}{2u}\Omega_m\Omega_d[-(\omega_m-\omega_d)^2 +\frac{Q}{3H}(\frac{1}{\rho_d}+\frac{1}{\rho_m})(\omega_m-\omega_d)]
\end{equation}\\

As $\omega_m$, the equation of state parameter for DM is always positive and decreases (or remains constant) with the evolution of the universe, so $\frac{d\omega_m}{du} \leq 0$. On the otherhand, the DE equation of state parameter $\omega_d$ is always negative and $\frac{d\omega_d}{du}$ can have any sign (i.e., +ve or -ve). Thus $\frac{dq}{du}$ may change sign more than once during the evolution of the universe and hence it is possible to have more than one real solution of q(u)=0. Therefore, it is possible to have more than one transition from deceleration to acceleration and vice versa in course of the evolution of the universe. However, in absence of interaction (i.e., Q = 0) if $\frac{d\omega_d}{du} \leq 0$ then q is a strictly decreasing function of u [34] and q(u) = 0 has exactly one real solution. Hence the interaction term has a significant effect in the transition from decelerating phase to accelerating phase or otherwise, although q has no explicit dependence on the interaction term Q.\\

\section{Cosmological models: Some examples}
For flat model, from the equation (21), q = 0 gives
\begin{equation}
\Omega_m\omega_m+\Omega_d\omega_d +\frac{1}{3}   = 0\\
~~~~~~~~~~~~~~~~~~~~~~~i.e,\omega_T +\frac{1}{3} = 0
\end{equation}\\

Also from the above equation we have
\begin{equation}
\omega_d = -\frac{\Omega_m\omega_m+\frac{1}{3}}{1-\Omega_m}
\end{equation} 
we shall now consider the following cosmological models and examine whether more than one transition is possible or not.\\
\subsection{Modified chaplygin gas:}
From the aspects of cosmological scenarios, the modified chaplygin gas (MCG) corresponds to radiation dominated universe at very early epochs and in the late epoch, the energy density behaves as cosmological constant, i.e., corresponds to the de Sitter universe. Moreover, due to much attention [35-37] of a unified description of CDM and DE in recent past, it is reasonable to use an exotic fluid namely modified chaplygin gas (MCG) model which describes the evolution of the universe from radiation era (by proper choice of the parameters) to $\Lambda CDM$. The equation of state for modified chaplygin gas (MCG) is given by [38]
\begin{equation}
p = \gamma\rho-\frac{B}{\rho^n}
\end{equation}\\
where $0 < \gamma \leq 1$ and $B, n > 0$. Now, if MCG is considered as the combined DM and DE then the equation of state parameter is given by
\begin{equation}
\omega_T = \frac{p}{\rho} = \gamma -\frac{B}{\rho^{n+1}}.
\end{equation}\\
So from equation (34) we have
\begin{equation}
\rho^{n+1} = \frac{B}{\gamma+\frac{1}{3}},
\end{equation}

which has only one feasible solution. Thus, there is only one transition from deceleration to acceleration as predicted by observations in recent past. Further, from the combined energy conservation equation, i.e., from $\dot{\rho} +3H(p+\rho) = 0$, for MCG model, we have on integration\\

~~~~~~~~~~~~~~~~~~~~~~~~~~$\rho^{n+1} =\frac{1}{\gamma+1}[C+\frac{B}{a^\mu}]$\\

where, $\mu$ =3($\gamma$+1)(n+1) and C is the constant of integration.
Hence the scale factor at the transition point is given by
\begin{equation}
a = [\frac{B(\gamma+\frac{1}{3})}{B(\gamma+1)-C(\gamma+\frac{1}{3})}]^\frac{1}{\mu}
\end{equation}\\
provided the constant of integration C is restricted by the relation $C < \frac{B(\gamma+1)}{\gamma+\frac{1}{3}}$.\\

Moreover, if MCG is chosen as the DE component then using the equation (35) we have
\begin{equation}
\rho^{n+1} = \frac{B}{\gamma+\frac{\Omega_m\omega_m+\frac{1}{3}}{1-\Omega_m}}
\end{equation}\\

Thus in any case we have only one transition from deceleration to acceleration to match with the present observation.\\
\subsection{Holographic DE model:}
If the DE is chosen in the form of holographic DE with equation of state parameter as equation (13) then from equation (35) we have\\

$\frac{1}{3}(1+\frac{2\sqrt{\Omega_d}}{c})+\frac{Q}{3H\rho_d} = \frac{\omega_m(1-\Omega_d)+\frac{1}{3}}{\Omega_d}$\\
This simplifies to the cubic equation (choosing x = $\sqrt{\Omega_d}$ )
\begin{equation}
\frac{2}{3c}x^3 +(\frac{1}{3}+\omega_m)x^2 +(b^2-\omega_m-\frac{1}{3}) = 0
\end{equation}\\
when Q=$3b^2H(\rho_m+\rho_d)$.\\

While we have the cubic equation 
\begin{equation}
\frac{2}{3c}x^3+(\lambda+\frac{1}{3}+\omega_m)x^2-(\omega_m+\frac{1}{3}) = 0
\end{equation}\\
when Q = $ 3\lambda H \rho_d $.\\

It should be noted here that both the cubic equations will have exactly one positive root (provided $ b^2 < \omega_m+\frac{1}{3}$) irrespective of whether $\omega_m = 0$ or not. So, in this case also we see that only one transition is possible.
\subsection{The phenomenological model of the form p = $-\rho-\frac{AB\rho^{2\alpha-1}}{A\rho^{\alpha-1}+B}$:}
This is a phenomenological model in which singularities can appear in the past or in future, depending on the choice of the parameters 'A' and 'B'. The following table shows the restrictions on the parameters of the model and the physical quantities that diverge, while the other quantities remain finite [39, 40].\\
$~~~~~~~~~~~~~~~~~~~~~~~~~~~~~~~~~~~~~~~~~~~~~~~~~~~~~~~~~${\bf Table I}\\\\
\begin{tabular}{|c|c|c|}
\hline Singularity & Divergences of the physical parameters & Restrictions on the model parameters\\
\hline Type I      & $a \rightarrow \infty$, $\rho \rightarrow \infty$, $p \rightarrow \infty$ & $\frac{3}{4} < \alpha < 1 $, $\forall$ A, B\\ 
\hline Type II     & $p\rightarrow \infty$                                                     & $\alpha < 0$ or $\frac{A}{B} > 1$\\
\hline Type III    & $\rho \rightarrow \infty$, $p \rightarrow \infty$                        & $\alpha > 1$ $\forall$ A, B\\
\hline Type IV     & $p \rightarrow \infty$                                                  & $0 < \alpha < \frac{1}{2}$, $\forall$ A, B\\
\hline
\end{tabular}\\

Note that for $\alpha = 1$, we have linear equation of state $p = \omega\rho$ with $\omega = -1-\frac{AB}{A+B}$, which is basically a constant and it is not of much interest. However, if $B < 0$ then strong energy condition, weak energy condition and null energy condition are satisfied but the dominant energy condition is violated here. Further, if $\alpha > 1$ then we see that\\
p $\rightarrow -\rho-A\rho^{2\alpha-1}$ as $\rho \rightarrow 0$\\
and p $\rightarrow -\rho-B\rho^\alpha$  as $\rho \rightarrow \infty$\\

Thus $\omega = \frac{p}{\rho} \rightarrow -1-0 (or -1+0)$ as $\rho \rightarrow 0$ and $A > 0 (or < 0)$, while $\omega \rightarrow +\infty (-\infty)$ when $\rho \rightarrow \infty$ and $B < 0 (or > 0)$.\\

If the DE component is chosen as the above fluid then the equation of state parameter takes the form
\begin{equation}
\omega_d = -1-\frac{AB\rho_d^{2\alpha-2}}{A\rho_d^{\alpha-1}+B}
\end{equation}
and from the equation (35) we have the quadratic equation
\begin{equation}
ABx^2 -A\delta x -B\delta = 0
\end{equation}\\
where $x = \rho_d^{\alpha-1}$ and $\delta = \frac{\Omega_m(1+\omega_m)-\frac{2}{3}}{1-\Omega_m}$\\
This quadratic equation has two positive real roots provided\\
{\bf I.} $B < 0$\\
{\bf II.} $\Omega_m < minimum [\frac{2}{3(1+\omega_m)}, \frac{\frac{2}{3}-4\frac{B^2}{A}}{1+\omega_m-\frac{4B^2}{A}}]$\\
and\\
{\bf III.} $\frac{4B^2}{A} < minimum [\frac{2}{3}, (1+\omega_m)]$\\

Thus under the above restrictions, it is possible to have two transitions for q(u) as shown in Figure 3. Further, as in the present model $\omega_m \geq 0$, restriction (III) implies $B^2 < \frac{A}{6}$. Moreover, if the DM is chosen in the form of dust, i.e., $\omega_m = 0$ and $B^2$ is chosen to be $A/8$, then the restriction (II) becomes $\Omega_m < minimum [\frac{2}{3}, \frac{1}{3}] = \frac{1}{3}$, which is well within the observed estimation of DM component.\\
$~~~~~~~~~~~~~~~~~~~~~~~~~~~$\includegraphics[height=4 in, width=4 in]{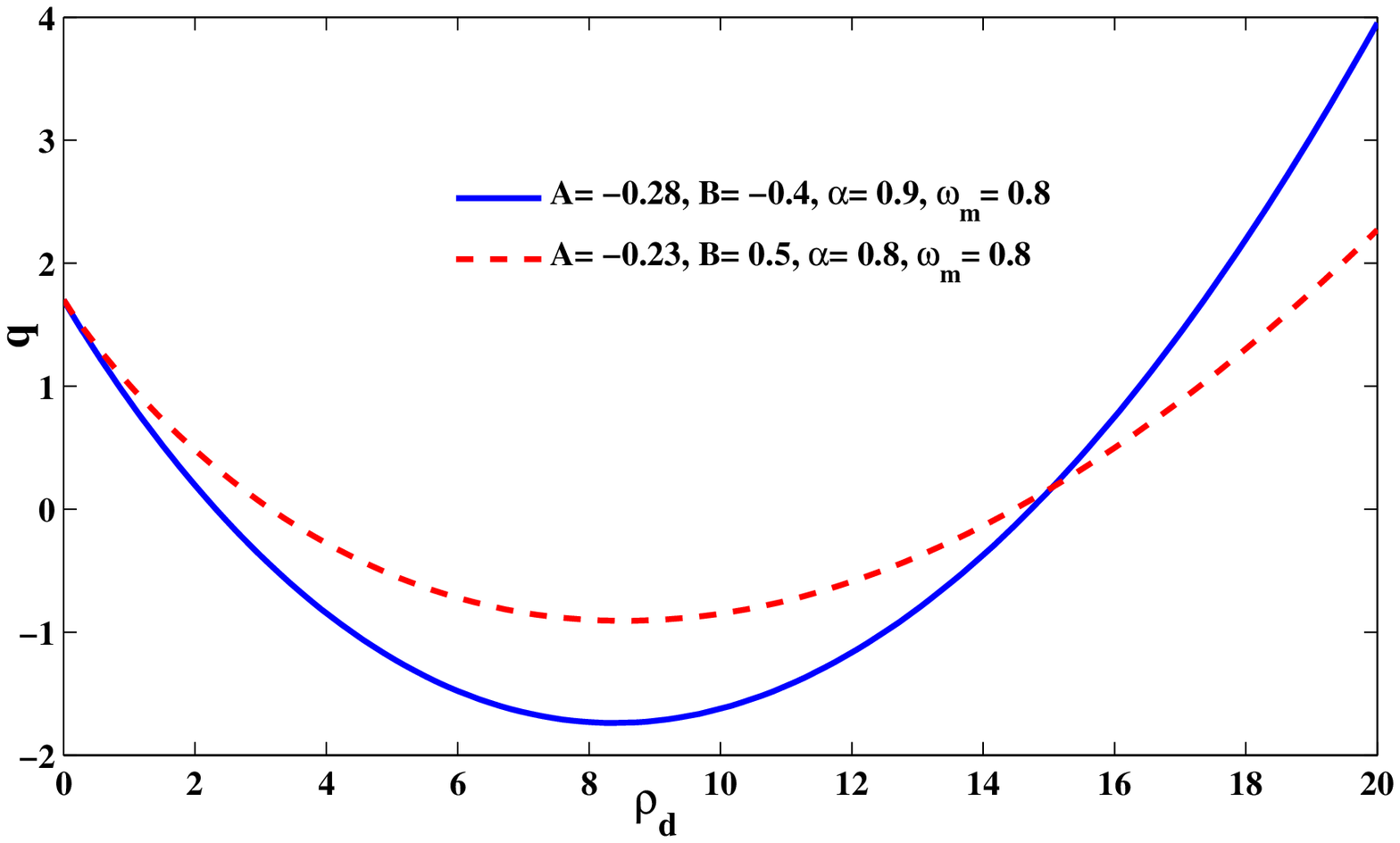}\\
Figure 3: These graphs show the variation of the deceleration parameter (q) with the dark energy density ($\rho_d$) for the phenomenological model described in 5.3.
\section{Discussion and concluding remarks:}
The present work is a model of interacting dark matter and dark energy where both the matter components are in the form of perfect fluid having barotropic equation of state. If due to interaction there is a energy flow from DE to DM then the effective equation of state parameters for both the fluids change significantly- DM has a tendency to become exotic while the effective DE may be still in quintessence era even when we have phantom DE. The evolution equations for energy densities of the dark species from a self-autonomous system provided the equation of state parameter for DM is taken as constant. We have a degenerate line of critical points which are non-hyperbolic in nature and hence the stability criteria can not be determined by linearization technique.\\
In the subsequent section we have evaluated a general expression for the decelertion parameter in terms of the density parameters and the equation of state parameters-it does not depend on the interaction term explicitly. However, the evolution of the deceleration parameter depends on the interaction term as well as on the evolution of the state parameters. As from physical consideration, the DM equation of state parameter $\omega_m$ should be positive but decreases with the evolution of the universe, while DE equation of state parameter $\omega_d$ is chosen as negative. So depending on the nature of the interaction term there may be more than one sign change in the expression $\frac{dq}{du}$ in course of the evolution of the universe. Thus, it is natural to speculate that the universe will again be in decelerating phase after the present accelerating era. In section 5, this speculation is examined for three different models. The modified chaplygin gas which is a unified DM and DE model has been investigated in subsection 5.1 and it is found that the model agrees with supernova observation, i.e., there is a smooth transition from deceleration to acceleration, but no future transition is predicted by the model. The interacting HDE model in subsection 5.2 also reveals similar conclusion. However, interesting feature is obtained in subsection 5.3 for the phenomenological model which in some sense behaves as DE. Here, in addition to the transition from deceleration to acceleration there is another possible transition from acceleration to deceleration as predicted by equation (33). Only future observations may indicate whether such future transition is possible or not.\\

Moreover, it should be noted that nine year WMAP data [41] and European-led research team behind the Planck cosmology probe show all-sky map of the cosmic microwave background (CMB) and according to them the present accelerating phase is due to the dominant DE component (68.37$\%$). So our models of DE evolution are well in  accord with recent observations. Further, the observations of CMB [42] anisotropies indicate the flatness of the universe.\\

Finally, for future work it will be interesting to examine the viability of the present DE models from the observational evidences particularly time (age-z) and distance (CMB/BAO) data.\\

\textbf{\large{Acknowledgements:}} Author SC is thankful to the UGC-DRS programme and author SP is thankful to The Council of Scientific and Industrial Research (CSIR), Govt. of India for Research grants.

\end{document}